\documentclass[twocolumn,superscriptaddress,floatfix,nofootinbib]{revtex4-2}

\usepackage{graphicx,graphics}
\usepackage{dcolumn}
\usepackage{amsmath,amssymb,amsfonts, amsthm}
\usepackage{latexsym,verbatim}
\usepackage{bm}
\usepackage{breqn}

\makeatletter
\let\cat@comma@active\@empty
\makeatother

\usepackage[toc,page]{appendix}
\usepackage{color}
\usepackage{ulem}
\usepackage{tikz}
\usepackage[breaklinks=true,colorlinks,citecolor=blue,linkcolor=blue,urlcolor=blue]{hyperref}
\usepackage{tabularx}
\usepackage{array, multirow, bigdelim, makecell, booktabs} 
\usepackage{braket}

\newtheorem{theorem}{Theorem}
\newtheorem{corollary}{Corollary}[theorem]
\newtheorem{definition}{Definition}

\graphicspath{fig}

\begin{document}

\title{Quantum approximation algorithms for many-body and electronic structure problems}

\author{Karen J. Morenz Korol}
\email{karen.morenz@mail.utoronto.ca}
\affiliation{Department of Chemistry, University of Toronto, Canada}
\affiliation{IBM Quantum, IBM T. J. Watson Research Center, Yorktown Heights, NY 10598, USA}
\author{Kenny Choo}
\email{kch@zurich.ibm.com}
\affiliation{IBM Quantum, IBM Research Zurich, Saumerstrasse 4, 8803 Ruschlikon, Switzerland}
\author{Antonio Mezzacapo}
\email{mezzacapo@ibm.com}
\affiliation{IBM Quantum, IBM T. J. Watson Research Center, Yorktown Heights, NY 10598, USA}


\begin{abstract}
Computing many-body ground state energies and resolving electronic structure calculations are fundamental problems for fields such as quantum chemistry or condensed matter. Several quantum computing algorithms that address these problems exist, although it is often challenging to establish rigorous bounds on their performances. Here we detail three algorithms that produce approximate ground states for many-body and electronic structure problems, generalizing some previously known results for 2-local Hamiltonians. Each method comes with asymptotic bounds on the energies produced. The first one produces a separable state which improves on random product states. We test it on a spinless Hubbard model, validating numerically the theoretical result. The other two algorithms produce entangled states via  shallow or deep circuits, improving on the energies of given initial states. We demonstrate their performance via numerical experiments on a 2-dimensional Hubbard model, starting from a checkerboard product state, as well as on some chemistry Hamiltonians, using the Hartree-Fock state as reference. In both cases, we show that the approximate energies produced are close to the exact ones. These algorithms provide a way to systematically improve the estimation of ground state energies and can be used stand-alone or in conjunction with existing quantum algorithms for ground states.

\end{abstract}

\maketitle

\section{Introduction}

Computing ground state energies of many-body systems or solving electronic structures for molecules and materials are critical problems appearing both in fundamental science and industrial applications. Although there is no known classical or quantum algorithm that can solve these problems exactly in the most general case~\cite{kempe2006complexity}, quantum computers have an exponential memory advantage over classical ones at attempting this task. The extent to which quantum computers can efficiently estimate the ground state of a many-body system, and the optimal algorithm for doing so, is an open area of research. 

Several quantum computing algorithms have been proposed for the ground state electronic structure problem. Phase estimation~\cite{kitaev1995quantum} requires long circuit depths and the preparation of an initial state with good overlap with the target ground state. The adiabatic algorithm~\cite{farhi2000quantum} can prepare ground states efficiently if the ground state gap in the adiabatic passage does not close. Variational quantum eigensolvers~\cite{peruzzo2014variational} search for ground states by optimizing gate angles of a quantum circuit used to prepare variational states on a quantum computer. They can perform well on shallow circuits~\cite{kandala2017hardware} amenable to be executed on noisy quantum devices. However, VQE algorithms can be unwieldy if the number of parameters to be optimized is too large - or conversely, if the number of parameters is too small, they may not be able to obtain a good estimate of the ground state energy. Recently, approaches based on imaginary-time evolution~\cite{motta2020determining} have been proposed, which rely on the assumption that correlations in the quantum state remain short-range as it undergoes the simulated imaginary dynamics.

Here we present an alternative approach, implementing quantum algorithms that can prepare approximate quantum ground states for many-body Hamiltonians and electronic structure problems. These algorithms build quantum circuits that take as input a target Hamiltonian and an initial quantum state, and efficiently construct a quantum circuit that produces a quantum state with lower energy. These quantum circuits have at most one variational parameter, independent of the size of the system considered. The results build on the theoretical results introduced in Ref.~\cite{Anshu2021} for two-local and k-local Hamiltonians, which in turn extended the results of~\cite{anshu2020beyond} for the Heisenberg model. We use these methods to target the electronic structure problem, and extend the results to deal with generic many body $k$-local Hamiltonians. These methods also allow to us construct approximate ground states for fermionic systems that go beyond the Gaussian states considered in~\cite{bravyi2019approximation}.  

These results detail three methods to build quantum circuits to get approximate ground states for $k$-local Hamiltonians, and we apply these algorithms to example electronic structure problems. The results are presented in order of increasing circuit depth, which give increasingly better approximations, as depth increases. We use numerical experiments to benchmark  the quality of the approximations on condensed matter and chemistry systems.

We start by defining the class of problems addressed. A $k$-local Hamiltonian can be written as: 
\begin{equation}
\label{eq: general H}
H =\sum_R h_R
\end{equation}
were each $h_R$ acts non-trivially at most on $k$ qubits, and thus can be written as a sum of (up to $4^k$) Pauli words, $h_R =\sum_{j=1}^{4^k} \gamma_j \boldsymbol{\sigma}_j$,
where $\gamma_j$ are real coefficients and $\boldsymbol{\sigma}_j$ are Pauli words:
$N$-fold tensor products of single-qubit Pauli operators $I, \sigma^x,\sigma^y,\sigma^z$, where at most $k$ of the single-qubit Pauli operators are non-identity. We think of this problem in terms of the hypergraph $\mathcal{G}(\mathcal{V}, \mathcal{E})$ defined by the Hamiltonian (\ref{eq: general H}), where the vertices $\mathcal{V}$ are the set of qubits, and each hyperedge $R \in \mathcal{E}$ connects the qubits corresponding to the non-identity elements in the term $h_R$. Within this graph representation the Hamiltonian is $k$-local when any hyperedge connects at most $k$ qubits. The degree $\deg(i)$ of a given qubit $i$  is defined as the number of hyperedges containing $i$, and we define $d=\max_{i\in V}(\deg(i))$. 

The second quantized fermionic Hamiltonian is a physically relevant subclass of the Hamiltonians defined by Eq.~(\ref{eq: general H}). They are usually formulated as
\begin{equation}
\label{fer_Ham}
H=\sum_{i, j} t_{ij} \, c^\dag_{i} c_{j} +  \sum_{i, j, k, m} u_{ijkm}\, c^\dag_{i} c^\dag_{k} c_{m} c_{j},
\end{equation}
where $c_i(c^\dag_i)$ are the $N$ fermionic annihilation (creation) operators for the mode i, satisfying the anticommutation relations $\{c^\dag_i, c_j\} = \delta_{i,j}$. The model in Eq.~(\ref{fer_Ham}) can be used to describe
the electronic structure problem in molecules and materials. It can be mapped to a $N$-qubit Hamiltonian of the type Eq.~(\ref{eq: general H}). The locality $k$ of the resulting mapped Hamiltonian depends on the specific encoding used. For example, we have $k=O(N)$ for the Jordan-Wigner mapping~\cite{Wigner1928}, $k=O(\log(N))$ if one uses a Bravyi-Kitaev mapping~\cite{bravyikitaev2002}, or $k=O(1)$ using superfast fermionic mappings~\cite{bravyikitaev2002, setia2018superfast} if the corresponding fermionic interaction graph is local. Here we shall use the Bravyi-Kitaev mapping in order to minimize locality while preserving generality.

\begin{figure*}[t]
    \centering
    \includegraphics[width=0.9\textwidth]{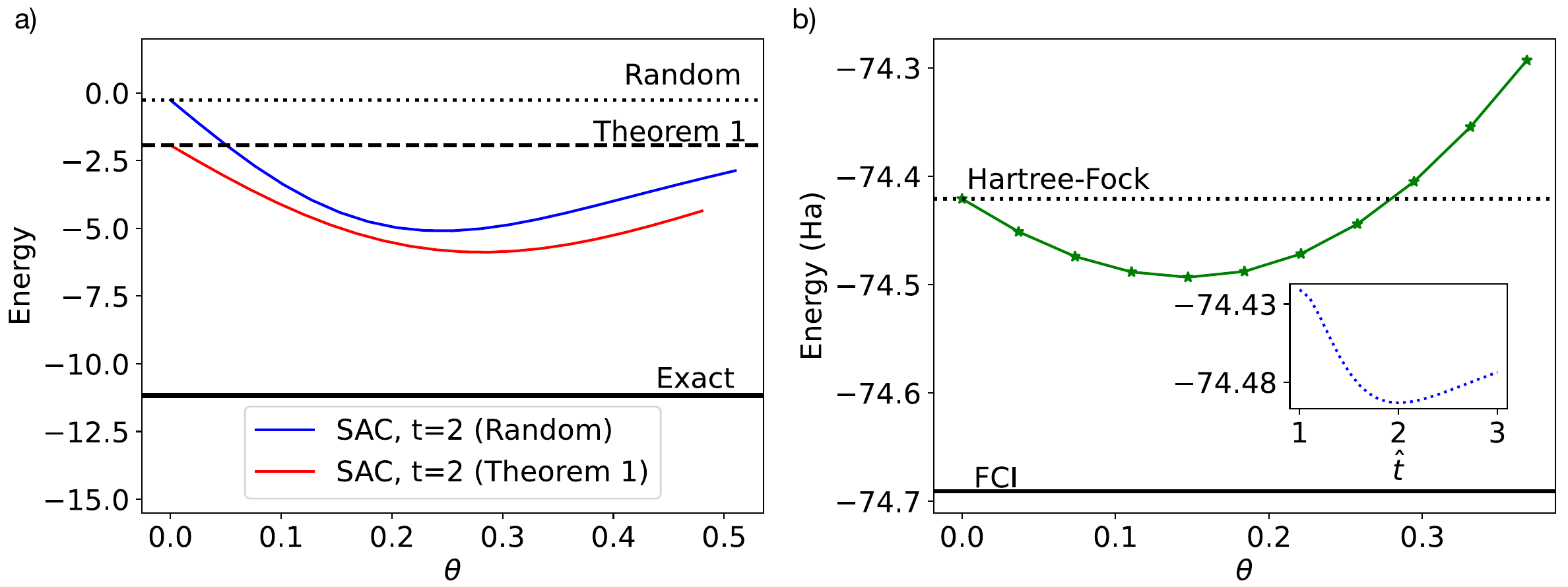}
    \caption{(a) SAC applied to the two-dimensional spinless Hubbard model $H = -\sum_{<ij>}t_i (c^\dag_ic_j + c^\dag_jc_i) +  \sum_{<ij>}v_{ij} c_i^\dag c_ic^\dag_jc_j$ with spatially disordered interaction strengths and nearest neighbor interactions. We compare the result starting from an arbitrary random state, with that of the state produced using the approach in Theoream 1. (b) SAC applied to a 20-qubit model of the $C_2$ molecule. We start from the Hartree-Fock state and choose $t=2$. The main plot shows energy of the state $\ket{\phi} = V(\theta)\ket{\textrm{HF}}$ as the parameter $\theta$ is varied. The dotted line indicates the energy of the Hartree-Fock state and the solid line gives the exact full configuration interaction (FCI) energy. In the inset, we show how the optimal energy varies when the single qubit operator is replaced according to Eq.~(\ref{eq: single qubit P}), with varying values of $t$. This shows that the energy is minimized when $\hat{t} = t$ as detailed in the SAC algorithm. } 
    \label{fig: C2 SAC}
\end{figure*}

\section{Approximation circuits for product states} 

Our first result is an extension to k-local Hamiltonians of a result for 2-local Hamiltonians in~\cite{Anshu2021}, which gives a circuit that produces a product state which outperforms the random state. In order to extend this result, we shall need the following two definitions to accommodate $k$-local Hamiltonians. 
\begin{definition}
A hypergraph $\mathcal{G}(\mathcal{V}, \mathcal{E})$ is triangle free if for any hyperedge $R_0$ containing two vertices $i,j\in\mathcal{V}$ there is no vertex $k\in\mathcal{V}, k\neq{i,j}$ that is connected both with $i$ through a hyperedge $R_1$ and with $j$ through a hyperedge $R_2$, where $R_0\neq R_1\neq R_2$. 
\end{definition}

We also need to define a function $f(O)$:

\begin{definition}
For an $n$-qubit $k$-local operator $O$, define $f(O)$ as the 2-norm of the $k$-local terms in the Pauli expansion of the operator, i.e. write
\begin{equation}
    O = \sum_{\vec{v}} \sum_{\vec{p}} \gamma^{\vec{v}}_{\vec{p}}\bigotimes_{i=1}^k \sigma_{p_i}^{v_i}
\end{equation}
where $\vec{v}$ is a list of qubits in strictly ascending order and $\vec{p}$ is a list of integers from $\{0,1,2,3\}$ pertaining to Pauli $I, X, Y, Z$. The  function $f$ is here defined as the square of the coefficients $\gamma$ for all strictly k-local terms, i.e. terms for which no qubit is acted upon by identity:
\begin{equation}
    f(O) = \sum_{\vec{v}} \sum_{\{\vec{p}: \forall i, \sigma_{p_i} \neq I\}} (\gamma^{\vec{v}}_{\vec{p}})^2\label{f}
\end{equation}
\end{definition}

We are now ready to state the following theorem

\begin{theorem}
Given a $k$-local Hamiltonian defined on a hypergraph $\mathcal{G}(\mathcal{V},\mathcal{E})$ which is triangle-free, it is possible to efficiently compute product states $\ket{v}$ which in expectation value satisfy
\begin{equation}
\begin{split}
\mathbb{E}_{v}\Big[\langle v | H | v \rangle\Big] \leq \mathrm{Tr}(H)/2^k - \Omega\Big(\dfrac{ f(H)}{\sqrt{d}3^{O(k)}2^{O(k)}}\Big)\label{product_bound}
\end{split}
\end{equation}
\end{theorem}

An inspection of Eq.~(\ref{f}) and the bound Eq.~(\ref{product_bound}) tells us that, for two Hamiltonians with the same number of terms, the improvement in energy would be greater for the Hamiltonian that is closer to being strictly $k$-local. This can be the case for example in condensed matter models defined on a lattice, or fermionic on lattices models mapped with superfast transformations~\cite{setia2018superfast} which preserve locality in the Pauli basis.

If we consider the Bravyi-Kitaev mapping for fermionic Hamiltonians as in Eq.~(\ref{fer_Ham}), where $k=O(\log(N))$, our bound becomes: 
\begin{equation}
\begin{split}
    \mathbb{E}[\langle v | H | v \rangle] \leq \mathrm{Tr}(H)/O(N) - \Omega\Big(\dfrac{ f(H)}{\sqrt{d}O(N)}\Big)\label{chem_product_bound},
\end{split}
\end{equation}
Where $N$ here represents the number of fermionic modes.

\section{Shallow approximation circuits} 

Shallow approximation circuits (SAC) can be built for the problems considered here using the approach described in~\cite{Anshu2021}, which lowers the energy of a given product state $\ket{v}$ by an amount proportional to the variance of that state with respect to the Hamiltonian. Note that although Ref.~\cite{Anshu2021} considers the problem of finding the maximum eigenvalue, this problem is equivalent to searching for the minimum eigenvalue (or ground state), since $-\lambda_{\min}(H) = \lambda_{\max}(-H)$. 

These circuits can be simulated efficiently on both a classical computer as well as a quantum device, so to obtain quantum advantage they could serve as a starting point for variational algorithms or phase estimation, or used in conjunction with higher-depth approximation circuits that will be presented later. 

Based on~\cite{Anshu2021}, given a $k$-local Hamiltonian as in Eq.~(\ref{eq: general H}), defined on a hypergraph $\mathcal{G}(\mathcal{V},\mathcal{E})$ of degree $d$, and an initial product state $\ket{v} := W\ket{0}$ (for some circuit $W$ which produces a product state), we can efficiently construct a circuit $U_S$ of depth $(d+1)$, such that the state $\ket{\psi_S} = U_S\ket{v}$ satisfies 
\begin{align}
    \langle \psi_S | H |\psi_S \rangle \leq \langle v | H |v \rangle - \Omega\left(\dfrac{Var_v(H)^2}{2^{O(k)}d^4 |\mathcal{E}|}\right)\label{SAC_bound}
\end{align}
where $Var_v(H)=\langle v|H^2|v\rangle -\langle v|H|v\rangle^2$ is the variance of $H$ with respect to $|v\rangle$. In the case of fermionic Hamiltonians, where $k=O(\log(N))$, we therefore have:

\begin{equation}
    \langle \psi_S | H |\psi_S \rangle \leq \langle v | H |v \rangle - \Omega\left(\dfrac{Var_v(H)^2}{O(N)d^4 |\mathcal{E}|}\right)
\end{equation}

We now give the procedure to construct the unitary $U_S$ that satisfies Eq.~(\ref{SAC_bound}).

\begin{enumerate}

    \item Let $S$ be the set that contains all collections $s$ of $t$ vertices $s = \lbrace j_1, \dots, j_t\rbrace$, for which $s$ is fully contained in the support of at least one hyperedge in $\mathcal{E}$. Let $t = \hat{t}$, where $\hat{t}$ is the value for which $\langle v|HQ_tH|v\rangle$ is maximized, and $Q_t$ is the projection onto weight $t$ operators for $0 \leq t \leq k$. 
    
    \item Define a single qubit operator acting on the i-th qubit as
    \begin{equation}
    \label{eq: single qubit P}
        P_i = 
        \begin{pmatrix}
        0 & \exp(-i\frac{\pi}{2t}) \\
        \exp(i\frac{\pi}{2t}) & 0
        \end{pmatrix}.
    \end{equation}

    \item For each collection $s \in S$,
    \begin{itemize}
        \item Define $P_s^P = P_{j_1} \dots P_{j_t}$,\newline and $P_s^X = X_{j_1}\dots X_{j_t}$
        \item Compute the commutator 
        \[
        \beta_{s}^p = \bra{v} [WP_s^pW, H] \ket{v}
        \] 
        for $p=P$ and $p=X$
        \item Define $a_s^p = \textrm{sign}(i \beta_{s}^p)$
    \end{itemize}
    \item Construct the Hermitian operator
    \begin{equation}
    \label{eq: SAC L}
        L^p = \sum_{s \in S} (-1)^{a_s^p} P_s^p
    \end{equation}
    \item Define the unitary operator
    \begin{equation}
    \label{eq: SAC}
        U_S^p(\theta) = e^{i\theta^p L^p}
    \end{equation}
    where $\theta^p = O\left( \frac{\beta^p}{k^2 d^2 \binom{k}{t-1}^2} \right)$ and $\beta^p = \sum_{s} |\beta_s^p|$. \newline $\theta^p$ can be optimized as a variational parameter, then $U_S^p \equiv U_S^p(\theta^*)$, where $\theta^*$ is the optimal parameter.
\end{enumerate}

Then for one or the other choice of $p$, $\ket{\psi_S}= U_S\ket{v}$ satisfies Eq.~(\ref{SAC_bound}).

Note that the unitary operator in Eq.~(\ref{eq: SAC}) can be implemented via Trotterization, and a single Trotter step can be performed with gate complexity $O(|S|)$. Also note that the circuit is shallow in the sense that the minimum circuit depth is independent of system size, and depends only on the maximum degree of any qubit in the graph. 

In general, this lower bound on the energy can be guaranteed only when implementing either the tensor product of the $P_i$ operators on every qubit or interchanging them with a tensor product of Pauli $X$ gates on every qubit, with the associated parameter $\theta$ \cite{Anshu2021}. However, in the setting of the electronic structure problem in Eq.~(\ref{fer_Ham}) with real coefficients and the Hartree-Fock state as input, the $P$ operator is always the best choice, for the following reason. The Hartree-Fock state is a computational basis state, prepared with a series of qubit $X$ gates, i.e. $\ket{v_{HF}} = W \ket{0}$ where $W = \bigotimes^\eta_{i=1}X_{p_i}$, and $\eta$ is the number of electrons in the system. Then both the terms $WP_s W$ and the Hamiltonian $H$ are Hermitian, so the commutator in step 3 is anti-Hermitian such that expectation values must be purely imaginary. On the other hand, both the Hartree-Fock state as well as the Hamiltonian are purely real for chemistry problems the term $WP_s^X W$ would be purely real and the expectation of the commutator would vanish. Thus, we need not consider the branch of the algorithm involving the $X$ gates in this case.

It should also be noted that both the expectation value of the commutator in step 3 as well as the optimization of the parameter $\theta$ in step 5 can be performed efficiently on a classical computer. 
In particular, we need to compute a maximum of $    {k \choose t} |E| 4^k$
commutator terms in order to calculate $\beta$, since $|S| = {k \choose t} |E| $ and the number of terms from the Hamiltonian required to calculate one of these commutator terms is upper bounded by $4^k$. For the electronic structure problem in Eq.~(\ref{fer_Ham}), using a Bravyi-Kitaev mapping~\cite{bravyikitaev2002}, it is possible to encode fermionic Hamiltonian such that the locality $k$ scales as $k = O(\log(N))$, so, using Stirling's approximation as $k$ (or $N$) goes to infinity, we can bound the size of $S$ as
\begin{equation*}
    \begin{split}
         O(|E| \log(N)^{-1/2})
    \end{split}
\end{equation*}
and therefore, the number of commutator terms we need to calculate as  
\begin{equation*}
    \begin{split}
         O(|E|N \log(N)^{-1/2})
    \end{split}
\end{equation*}
Thus, the scaling is sub-exponential in system size. 
Moreover, the circuit $U_S$ (\ref{eq: SAC}) can be implemented efficiently as a shallow circuit on a quantum device since the terms in the operator $L$ (\ref{eq: SAC L}) are mutually commuting.

\begin{figure*}[t]
    \centering
    \includegraphics[width=0.9\textwidth]{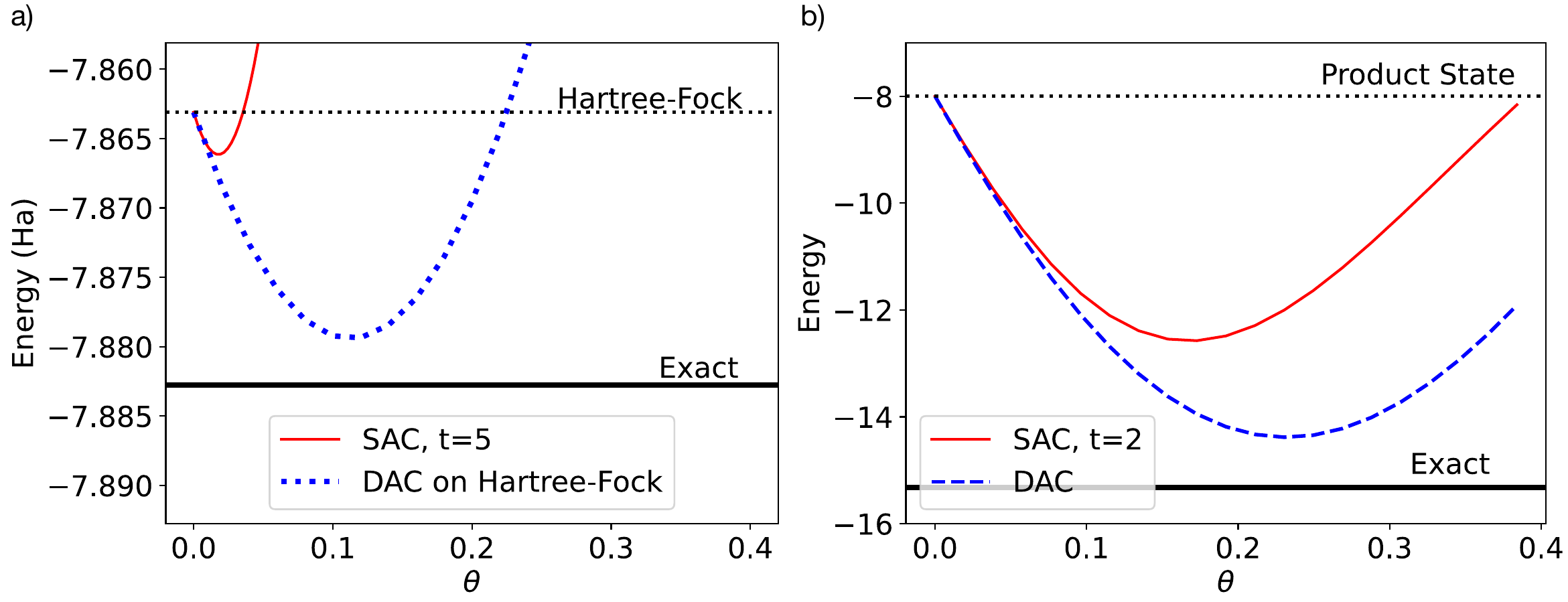}
    \caption{(a) Performance of the SAC and DAC applied to $LiH$ molecule in the STO-3G basis. (b) Performance of the SAC and DAC applied to the spinless Hubbard model on a $4$ by $4$ square lattice, with periodic boundary conditions and nearest neighbour interaction $U/t=1$.}
    \label{fig: DAC}
\end{figure*}

The results of numerical tests of the SAC approach are shown in Fig.~\ref{fig: C2 SAC}, for a two-dimensional spinless Hubbarb model, and for a molecular system: C$_2$ in the STO-3G basis. In the case of the two-dimensional Hubbard model, we consider a $4$x$4$ square lattice with spatially disordered interactions, in Fig.~\ref{fig: C2 SAC}a, initializing the product state in both a random product state as well as the state produced with the approach in Theorem 1. We show a sizable improvement in the approximate ground state energy. For the molecular simulation shown in Fig.~\ref{fig: C2 SAC}b, we start from the Hartree-Fock state, applied to the molecule $C_2$ in the STO-3G basis. In the inset of Fig.~\ref{fig: C2 SAC}b, we show that the choice of the $t$ parameter for the single qubit operator Eq.~(\ref{eq: single qubit P}) is optimal. We see in Fig.~\ref{fig: C2 SAC}b that the shallow circuit $U_S$ for the optimal value of $\theta$ is able to capture about $25\%$ of the correlation energy, improving on the Hartree-Fock energy.

\section{Deep approximation circuits}

We consider now a second class of approximation circuits, which can improve upon entangled initial states as well as product states, using quantum circuits of higher depths. To begin, we define lightcones:

\begin{definition}[Lightcone]
For any $n$-qubit quantum circuit and any qubit $j\in [n]$, we define the lightcone $\ell(j) \subseteq [n]$ as the set of all output qubits that are causally connected to $j$. 
\end{definition}

Then, we define the maximum lightcone size $\ell = max_{j\in [n]} |\ell(j)|$. Note that for any depth $d$ circuit composed of two-qubit gates, we have $\ell \leq 2^d$. 

We then extend a result proven in~\cite{Anshu2021} for 2-local systems to arbitrary $k$-local systems, summarized in the theorem here below. 
\begin{theorem}\label{Theorem:Deep Approximation Circuits}
Given a $k$-local Hamiltonian $H$ and a quantum state $\ket{\psi} = W \ket{0}$ generated by a unitary $W$ with maximum lightcone size $\ell$, it is possible to construct a state $|\psi_D\rangle = U_D|\psi\rangle$ such that 
\begin{equation}
    \langle \psi_D |H|\psi_D \rangle \leq \langle \psi |H|\psi \rangle - \Omega \left(\dfrac{Var(H)^2}{k^4\ell^{10}d^2|\mathcal{E}|}\right).
    \label{DAC_bound}
\end{equation}
\end{theorem}
This result follows directly from the proof of Theorem 3 in \cite{Anshu2021} by replacing the relevant factors of 2 (which come from considering the 2-local case) with $k$ for our $k$-local case. 

Unlike the shallow approximation circuits, the circuit $U_D$ can be applied to an arbitrary initial state; we do not require $\ket{\psi}$ to be a product state. 
The circuit $U_D$ satisfying Theorem~\ref{Theorem:Deep Approximation Circuits} is constructed starting from a $k$-local Hamiltonian $H$ as well as an initial state $\ket{\psi} = W \ket{0}$, as follows:
\begin{enumerate}
    \item Obtain the operator
    \begin{equation}
    \label{eq: DAC F}
        F = \sum_{j=1}^{N} W \ket{1}\bra{1}_j W^{\dagger},
    \end{equation}
    where $\ket{1}\bra{1}_j$ is the projector on the excited state of qubit $j$.
    \item Compute the commutator
    \begin{equation}
    \label{eq: DAC A}
        A = i[H, F]
    \end{equation}
    \item Then the deep approximation circuit is given by
    \begin{equation}
    \label{eq: DAC U}
        U_D = e^{-i\theta A}
    \end{equation}
\end{enumerate}
It should be noted the circuit $U_D$ (\ref{eq: DAC U}) can no longer be simulated efficiently on a classical computer, however, low-order trotter schemes \cite{lloyd1996universal} still allow for simulation on quantum devices with a resource requirement scaling polynomially in the number of Pauli words in the operator $A$ (\ref{eq: DAC A}). 

In order to calculate $A$, one needs to compute the commutator in Eq.~(\ref{eq: DAC A}), which corresponds to the set of commutators $[h_R,F]$. Each $h_R$ consists of a maximum of $4^k$ Pauli words. In the case where $W$ prepares the Hartree-Fock state, taking the commutator expands this by up to a factor of $k$, since the commutators where $j$-th qubit of Eq.~(\ref{eq: DAC F}) is not contained in the support of $h_R$ will be zero. Thus, overall we have to calculate
\begin{equation*}
     k4^k|E|
\end{equation*}
terms. In the case of fermionic Hamiltonians using the Bravyi-Kitaev mapping, where $k$ scales as $O(\log(N))$, this simplifies to:
\begin{equation*}
     O(N\log(N)|E|)
\end{equation*}
Note that, in the case of the fermionic Hamiltonians, the bound in Eq.~\ref{DAC_bound} becomes simply:
\begin{equation}
    \langle \psi_D |H|\psi_D \rangle \leq \langle \psi |H|\psi \rangle - \Omega \left(\dfrac{Var(H)^2}{O(\log(N)^4)\ell^{10}d^2|\mathcal{E}|}\right).
    \label{DAC_bound_fermion}
\end{equation}

In the case where the initial state $\ket{\psi}$ is the Hartree-Fock state, $W$ is again a single Pauli word such that the operator $F$ in Eq.~(\ref{eq: DAC F}) will consist of only $N$ Pauli words. It then follows that $A$ will be a sum of $O(N_{H} \times N)$ Pauli words where $N_H$ is the number of terms in the Hamiltonian. If a first order trotterization is used, simulation on a quantum computer would then have a gate complexity of $O(N_{H} \times N)$.

We test the performance of the deep approximation circuit for the LiH molecule in the STO-3G basis, defined on 12 qubits, and a spinless Hubbard model on a square, 4x4 lattice, with results shown in Fig.~\ref{fig: DAC}. We take $W$ to be the unitary that prepares the Hartree-Fock state for LiH. This shows that whereas the SAC only manages to capture about $20\%$ of the correlation energy, the DAC, which can also be seen as a variational algorithm with only one parameter, captures nearly $80\%$, reaching an energy that deviates from the exact by some milliHartrees. Similar results are observed for the spinless two-dimensional Hubbard model in Fig.~\ref{fig: DAC}b, where we have chosen $W$ such that it instantiates a checkerboard product state. Again, the DAC improves on the result obtained from the SAC.

\section{Conclusion}

We have implemented quantum approximation algorithms for $k$-local quantum Hamiltonians and electronic structure problems. These algorithms involve the construction of quantum circuits that can prepare good candidate ground states for the target Hamiltonians considered. A first class of these quantum circuits produces product states with a guaranteed upper bound on their expected energy which is lower than the expected energy achieved by the random state. This first approximation algorithm is relevant for condensed matter models for which a mean field approximation does not give good results, or as a seed for the other algorithms. 

The other approximation algorithms take as input an initial state, and construct a circuit to create a new entangled state which improves the energy estimate by an amount proportional to the variance of the initial state. In the first algorithm, only product states can be taken as input, but the circuit is always shallow, while the second more general algorithm can accept entangled states as input, but may not result in a shallow circuit. 

Thus, this paper demonstrates a step-wise procedure to first prepare a product state estimation of the ground state energy of a target Hamiltonian, which can then be improved upon by the shallow circuit. Then, it is possible to further improve the estimate by feeding this result into the deep approximation circuit, but it remains to be seen in which scenarios this can be computed efficiently. Instead, we compare the result of the SAC and the DAC on the same input state, and demonstrate that the DAC achieves a greater energy improvement. 

Finally, we give some considerations on how to use the results presented in the context of the existing quantum algorithms for ground state problems. The circuits here used to prepare approximate ground states can be seen as variational circuits that only use one parameter. Further research can explore the use of these circuits as starting states for variational quantum eigensolvers or phase estimation algorithms. In addition, these circuits could be used as blueprints to design better variational ansatzae that involve more variational parameters.

The numerical experiments performed show that the methods presented are effective for preparing approximate ground states of many-body and molecular systems of small systems, approaching the full CI solution for the LiH ground state. Future experiments on classical and quantum computers should assess whether the quality of the approximations is consistent on systems with larger sizes. We foresee the approaches presented here as novel tools to tackle the ground state problem of many-body $k$-local and electronic structure Hamiltonians on quantum computers.   

\section*{Acknowledgements}

We are grateful to Bryce Fuller and Charles Hadfield for fruitful discussions. KJMK acknowledges support from NSERC Vanier Canada Graduate Scholarship. 

\bibliography{notesbib}

\onecolumngrid

\newpage

\section*{Supplementary Information}

\subsection{Proof of Theorem 1}

We will prove this theorem in the convention of classical optimization, where we will look for the maximum eigenvalue. As noted in the main text, this problem is equivalent to searching for the minimum eigenvalue, since $-\lambda_{\min}(H) = \lambda_{\max}(-H)$. Let $G(V,E)$ be a hypergraph with hyperedges of size at most $k$ and at most $N=|V|$ qubits at the vertices. Denote $deg(i) \leq d$ for the number of hyperedges containing $i\in V$. Consider a k-local Hamiltonian $H=\sum_{R\in E} h_R$ where each $h_R$ acts on a subset $R \in E$ of qubits, $|R| \leq k$ and $\|h_R\|\leq 1$, where this is the spectral norm, defined as the maximum length of output vector due to the operation of the operator on any vector of length less than or equal to 1 (see e.g. \cite{watrous_2018}). Note that the expected energy of $H$ with respect to the random state $\rho = I/2^n$ is $\mathrm{Tr}(H)/2^n$.
Following the proof of Theorem 5 in \cite{Anshu2021}, we will work in a local Pauli basis chosen randomly and independently for each qubit, and write $h_R$ in this basis. 

Let $w_R = \mathrm{Tr}(h_R)/2^k$ and $u_R^x = \mathrm{Tr}(h_R \prod_{i \in R} X_i)/2^k$, and likewise for $u_R^y$ and $u_R^z$.
We pick a random i.i.d. assignment of pure product $|v\rangle = \bigotimes_{i=1}^n |v_i\rangle$ states to the vertices. We can represent the state of each qubit as $\rho_i = (I + r_i^x X_i + r_i^y Y_i + r_i^z Z_i)/2$ using the unit Bloch vector $\{r_i^x, r_i^y, r_i^z\}$. 

Now we select a subset of vertices $A$ uniformly at random, meaning that we include each vertex in $A$ with a probability $1/2$. Define a set $B_i$ as the set of hyperedges $R$ which have exactly one vertex $i \in A$. For any vertex $i\in A$ and $R \in B_i$, let $N_R(i) = \{j \in R: j \notin A\}$ 

With this framework, the total energy of a given hyperedge associated to $h_R$, with only one vertex $i \in A$ is:

\begin{equation}
    \mathrm{Tr}(h_R\bigotimes_{j\in R} \rho_j) = w_R + \sum_{a \in \{x, y, z\}} u_R^a \prod_{i\in R} r_i^a + D,
\end{equation}
where we have conveniently hidden all the cross terms (terms containing more than one type of Pauli operator) in this $D$. Then we can write the energy of all the hyperedges which contain no vertex in $A$ except $i$ as:
\begin{equation}
    \sum_{R \in B_i} \mathrm{Tr}(h_R\bigotimes_{j\in R} \rho_j) = 
    \sum_{R \in B_i}w_R  
     + \sum_{R \in B_i}\sum_{a \in \{x, y, z\}} u_R^a \prod_{j\in R} r_j^a 
     + \sum_{R \in B_i}D
\end{equation}
Now rotate qubit defined by the new Bloch vector (for $a\in \{x,y,z\}$):

\begin{equation}
    \tilde{r}_i^a = \sum_{R\in B_i} u_R^a \prod_{j\in R, j \neq i}\frac{r_j^a} {\sum_{b\in \{x,y,z\}} (\sum_{R\in B_i}  u_R^b \prod_{j\in R, j \neq i}r_j^b)^2}\label{tilde_r}
\end{equation}
Recall here that $u_R^b= \mathrm{Tr}(h_R \prod_{i \in R} X_i)/2^k \neq 0$ only when $h_R$ acts with the same Pauli operator on all the vertices sharing the hyperedge $R$. Only these types of interactions contribute to the denominator of Eq.~(\ref{tilde_r}). The qubit $i$ is therefore rotated into a state whose $\tilde{r}_i^a$ component 
Then, the second term in the energy of all the hyperedges which contain no vertex in $A$ except $i$ becomes:

\begin{equation}
\begin{split}
     \sum_{a \in \{x, y, z\}}\Big[&\Big(\sum_{R \in B_i} u_R^a \prod_{j\in R, j \neq i} r_j^a\Big)  \\
     &\Big(\sum_{R\in B_i} u_R^a \prod_{j\in R, j \neq i} r_j^a \Big)\Big]\\
     &\Big(\sum_{b\in \{x,y,z\}}(\sum_{R\in B_i} u_R^b \prod_{j\in R, j \neq i}  r_j^b)^2 \Big) ^{-1/2}\\
     &= \Big(\sum_{a \in \{x, y, z\}}(\sum_{R\in B_i} u_R^a \prod_{j\in R, j \neq i}  r_j^a)^2 \Big) ^{1/2}\\
\end{split}
\end{equation}

Now we want to think about the total energy. There are three kinds of hyperedges:

\begin{enumerate}
    \item Hyperedges which contain no vertices in A - call this set N
    \item Hyperedges with exactly one vertex in A
    \item Hyperedges with multiple vertices in A - call this set M
\end{enumerate}

Thus, we can write the overall expected energy of our updated state $\tilde{\rho}$ as:

\begin{equation}
\begin{split}
    \mathbb{E}(\mathrm{Tr}(H\tilde{\rho}))& = \mathbb{E}(\sum_{R\in N}\mathrm{tr}(h_R\bigotimes_{j\in R} \rho_j)) \\
&+ \mathbb{E}(\sum_{i\in A}\sum_{R \in B_i}\mathrm{tr}(h_R\bigotimes_{j\in R, j \neq i} \rho_j\otimes \tilde{\rho}_i))\\
&+ \mathbb{E}(\sum_{R\in M}\mathrm{tr}(h_R\bigotimes_{j\in R \cap A} \tilde{\rho}_j \bigotimes_{k\in R \cap A^c} \rho_k)) 
\end{split}    
\end{equation}

Note that $\mathbb{E}(\tilde{r}_i^a) = 0$ in all cases, and that $\tilde{r}_i, \tilde{r}_j$ are independent of each other whenever $i$ and $j$ are both vertices contained in a hyperedge $R\in M$. This follows from the definition of the set $M$, the initial uniform i.i.d. distribution of the state of vertices, and the triangle-freeness, which we define for hyperedges as follows. In this case, triangle-free means that for any
vertices $i,j$, if there is a hyperedge $R_{ij}$ containing both $i$ and $j$, then there is no vertex $k$ for which there is both a hyperedge $R_{ik}$ containing both $i$ and $k$, along with a hyperedge $R_{jk}$ containing both $j$ and $k$. The hyperedges $R_{ij}$, $R_{ik}$, and $R_{jk}$ would be considered a triangle. 

Furthermore, we can see that if $a \neq b$ then $\mathbb{E}(r_j^ar_k^b) = \mathbb{E}(\tilde{r}_j^ar_k^b) = 0$. If $a=b$, we still have $\mathbb{E}(r_j^ar_k^b) = 0$, but $\tilde{r}_j^a$ could depend on $r_k^a$, and both can appear in a single term in $D$ where for some other qubit we have $r_{\ell}^b$, $b \neq a$. But then, from the first part, the overall expectation value of such a term is still zero, and so our $D$ term disappears. 

Thus we can write:

\begin{equation}
    \begin{split}
    \mathbb{E}(\mathrm{Tr}(H\tilde{\rho}))& = \sum_{R\in E} w_R\\
    &\quad +  \mathbb{E}(\sum_{i\in A} (\sum_{a\in {x,y,z}} (\sum_{R\in B_i} u_R^a \prod_{j\in R; j\neq i} r_j^a)^2)^{1/2})\\
    & \geq \sum_{R\in E} w_R + \mathbb{E}(\sum_{i\in A} | \sum_{R\in B_i} u_R^z \prod_{j \in R; j\neq i} r_j^z|)
\end{split}   
\end{equation}

From this point on, the rest of the proof simply applies directly, when we replace $\xi_i$ with $ \sum_{R\in B_i} u_R^z \prod_{j \in R; j\neq i} r_j^z$. Following along:

The first term in equation (10) corresponds to the expected energy when the product states are chosen uniformly at random. The second term gives us some improvement achieved by the local updates, and so we need to figure out the bound arising from that. 

For a fixed choice of the set $A \subseteq V $, define the random variable $\xi_i = \sum_{R\in B_i} u_R^z\prod_{j\in R; j\neq i} r_j^z$. Using the second moment method for $t\in[0,1]$, 

\begin{equation}
    \mathrm{Pr}[|\xi_i|\geq t\sqrt{\mathbb{E}[\xi^2]}] \geq (1-t^2)^2 \dfrac{\mathbb{E}[\xi^2]^2}{\mathbb{E}[\xi^4]}
\end{equation}

Note that this follows from the Paley-Zygmund inequality, explained as follows. Suppose that $X_n$ is a sequence of non-negative real-valued random variables which converge in law to a random variable X. If there are finite positive constants $c_1$ and $c_2$ such that $\mathbb{E}[X_n^2] \leq c_1 \mathbb{E}[X_n]^2$ and $\mathbb{E}[X_n] \geq c_2$ for all $n$, then for every $n$ and $\theta \in (0,1)$, we have $\mathrm{Pr}(X_n \geq c_2 \theta) \geq (1-\theta)^2/c_1$. In order to obtain equation 11, we set $\theta = t^2$, $X_i = \xi_i^2$, $c_1 = \mathbb{E}[\xi_i^4]/(\mathbb{E}[\xi_i^2]^2)$, and $c_2 = \mathbb{E}[\xi_i^2]$. Then we note that we can take the square root of the terms inside the bracket on the lefthand side without changing anything. 

Now we will sample uniform pure states over the Bloch sphere by drawing $\phi \in [0,2\pi]$ and $r_j^z \in [-1,1]$, and then setting $r_j^x = \sqrt{1-(r_j^z)^2}\cos\phi$ and $r_j^x = \sqrt{1-(r_j^z)^2}\sin\phi$. Then, $\mathbb{E}[r_j^z] = 0$, $\mathbb{E}[(r_j^z)^2] = 1/3$, $\mathbb{E}[(r_j^z)^3] = 0$, and $\mathbb{E}[(r_j^z)^4] = 1/5$. 

Consider Corollary 9.6 of \cite{o2014booleananalysis}: 

\begin{corollary}
Let $x_1, ..., x_n$ be independent, not necessarily identically distributed, random variables satisfying $\mathbb{E}[x_i] = \mathbb{E}[x_i^3] = 0$ (which holds, for example, if each $-x_i$ has the same distribution as $x_i$.) Assume also that each $x_i$ is B-reasonable (defined below). Let $f = F(x_1, ..., x_n)$ where $F$ is a multilinear polynomial of degree at most $a$. Then $f$ is $\max(\mathrm{B},9)^a$-reasonable. 
\end{corollary}

\begin{definition}
 We say that a real random variable $X$ is $B$-reasonable if for a real number $B\geq 1$,  
\begin{equation}
    \mathbb{E}[X^4] \leq B\mathbb{E}[X^2]^2
\end{equation}
\end{definition}

Now we have for a fixed choice of set $A$, and a given vertex $i$, the expectation with respect to the random distribution of initial product states for some $t\in [0,1]$ is:

\begin{equation}
    \begin{split}
        \mathbb{E}\Big[|\sum_{R \in B_i} u_R^z \prod_{j\in R, j\neq i}r_j^z|\Big]\geq \dfrac{1}{9} t(1-t^2)^2 \sqrt{\mathbb{E}[\xi_i^2]}
    \end{split}
\end{equation}

Briefly consider the final term: 

\begin{equation}
    \begin{split}
        \mathbb{E}[\xi_i^2] &= \mathbb{E}\Big[(\sum_{R\in B_i}u_R\prod_{j\in r; j \neq i} r_j^z)^2\Big]\\
        &= \mathbb{E}\Big[(\sum_{R\in B_i}(u_R^z)^2\prod_{j\in r; j \neq i} (r_j^z)^2)\Big]
    \end{split}
\end{equation}

Since if $R, S \in B_i$, if $R\neq S$ then there exists a vertex $j\in R$, $j\notin S$. But, as above, $\mathbb{E}[r_j^z] = 0$, so any such term will disappear, and therefore we can ignore all terms with $R \neq S$. Thus, 

\begin{equation}
    \begin{split}
        \mathbb{E}\Big[|\sum_{R \in B_i} u_R^z \prod_{j\in R, j\neq i}r_j^z|\Big]\geq \dfrac{1}{9} &t(1-t^2)^2 \sqrt{\mathbb{E}[\xi_i^2]}\\
        = \dfrac{1}{9} &t(1-t^2)^2\cdot\\
        &\sqrt{\sum_{R\in B_i}(u_R^z)^2\prod_{j\in r; j \neq i} \mathbb{E}[(r_j^z)^2)]}\\
        \geq \dfrac{1}{9\cdot3^{(k-1)/2}} &t(1-t^2)^2\cdot\sqrt{\sum_{R\in B_i}(u_R^z)^2}\\
        = \dfrac{1}{3^{(k+3)/2}} &t(1-t^2)^2\cdot\sqrt{\sum_{R\in B_i}(u_R^z)^2}
    \end{split}
\end{equation}

We now calculate the expectation with respect to $A \subseteq V$. Note that $B_i$ is also a random variable determined by the set $A$. Now we consider Theorem 9.24 of \cite{o2014booleananalysis}:

\begin{theorem}
Let $f:\{-1,1\}^n\rightarrow \mathbb{R}$ be a nonconstant function of degree at most $a$. Then 
\begin{equation}
    \mathrm{Pr}_{x\in \{-1,1\}^n}[f(x) > \mathbb{E}[f]]\geq (4e^{2a})^{-1}
\end{equation}
\end{theorem}

Conditioned on the event that $i \in A$, we consider the random variables to be whether or not a vertex $j$ is in one of the $R \in B_i$ to calculate:

\begin{equation}
    \begin{split}
        \mathrm{Pr}\Big[ \sum_{R \in B_i} (u_R^z)^2\geq \dfrac{1}{2^{k-1}} \sum_{R \in E ; i\in R} (u_R^z)^2 \Big] \geq \dfrac{1}{4e^2}
    \end{split}
\end{equation}

Since we expect that $\dfrac{1}{2^{k-1}}$ edges containing vertex $i$ are in $B_i$, where each individual vertex has a 50-50 chance of being in $A$. Therefore we have:

\begin{equation}
    \begin{split}
        \mathbb{E}\Big[\sum_{i\in A}&(\sum_{R\in B_i}(u_R^z)^2)^{1/2}\Big] \\
        &\geq \dfrac{1}{4\cdot2\cdot2^{(k-1)/2} e^2}\sum_{i\in V}\Big(\sum_{R\in E; i\in R}(u_R^z)^2\Big)^{1/2}\\
        &=\dfrac{1}{2^{(k+5)/2} e^2}\sum_{i\in V}\Big(\sum_{R\in E; i\in R}(u_R^z)^2\Big)^{1/2}\dfrac{\sqrt{d}}{\sqrt{d}}\\
        &=\dfrac{1}{2^{(k+5)/2} e^2}\sum_{i\in V}\Big(\sum_{R\in E; i\in R}\dfrac{(u_R^z)^2}{d}\Big)^{1/2}{\sqrt{d}}\\
        &\geq\dfrac{1}{2^{(k+5)/2} e^2}\sum_{i\in V}\Big(\sum_{R\in E; i\in R}\dfrac{(u_R^z)^2}{d}\Big){\sqrt{d}}\\
        &\geq \dfrac{2}{2^{(k+5)/2} e^2 \sqrt{d}}\sum_{R\in E}(u_R^z)^2\\
        &= \dfrac{1}{2^{(k+3)/2} e^2 \sqrt{d}}\sum_{R\in E}(u_R^z)^2\\
    \end{split}
\end{equation}

Where we used the fact that $(u_R^z)^2 \leq 1$, and thus $\sum_{R\in E; i\in R}(u_R^z)^2 \leq d$, so that we can ignore the power of $1/2$, along with the fact that that there are at least $2$ vertices per hyperedge. 

Thus, taking the expectation over random basis, we arrive at:

\begin{equation}
    \begin{split}
        \mathbb{E}\Big[\sum_{i\in A}&(\sum_{R\in B_i}(u_R^z)^2)^{1/2}\Big] \\
        &\geq \dfrac{1}{3^k2^{(k+3)/2} e^2 \sqrt{d}}f(H)\\
    \end{split}
\end{equation}

where $f(H)$ is the 2-norm of the k-local terms in the Pauli expansion of the Hamiltonian, i.e. for an n-qubit k-local operator

\begin{equation}
    O = \sum_{\vec{x}} \sum_{\vec{p}} \gamma^{\vec{x}}_{\vec{p}}\bigotimes_{i=1}^k \sigma_{p_i}^{x_i}
\end{equation}

where $\vec{x}$ is a list of qubits in strictly ascending order and $\vec{p}$ is a list of integers from $\{0,1,2,3\}$ pertaining to Pauli $I, X, Y, Z$. Then we define the function $f$ as the square of the coefficients $\gamma$ for all k-local terms, i.e. terms for which no qubit is acted upon by identity:

\begin{equation}
    f(O) = \sum_{\vec{x}} \sum_{\{\vec{p}: \forall i, \sigma_{p_i} \neq I\}} (\gamma^{\vec{x}}_{\vec{p}})^2
\end{equation}

We plug this in to the above to get: 

\begin{equation}
    \begin{split}
        \mathbb{E}\Big[\sum_{i\in A}& |\sum_{R \in B_i} u_R^z \prod_{j\in R, j\neq i}r_j^z|\Big]\\
        &\geq \dfrac{1}{3^{(k+3)/2}} t(1-t^2)^2\cdot\Big(\dfrac{1}{3^k2^{(k+3)/2} e^2 \sqrt{d}}f(H)\Big)\\
        &= \dfrac{t(1-t^2)^2 f(H)}{27^{(k+1)/2}2^{(k+3)/2} e^2 \sqrt{d}} 
    \end{split}
\end{equation}

Thus, overall we have: 

\begin{equation}
    \begin{split}
        \mathbb{E}\Big[\langle \psi | H | \psi \rangle\Big] \geq \mathrm{Tr}(H)/2^k + \Omega\Big(\dfrac{ f(H)}{\sqrt{d}3^{O(k)}2^{O(k)}}\Big)
    \end{split}
\end{equation}

\subsection{Complexity for the shallow approximation circuits}

 Here we explain the classical computational overhead necessary to define the shallow approximation circuit defined in the main text. Borrowing notation from the main text, there are 
\begin{equation*}
    |S| = {k \choose t} |E|
\end{equation*}
sets of $\{j_1,...,j_t\}$ in $S$, and for each, we must compute a commutator of a simple operator (either a string of Pauli $X$ or the operator $p$ above) with $H$. By definition, each set of $\{j_1,...,j_t\}$ is contained fully in the support of one $h_R$, so for each of these $\{j_1,...,j_t\}$, 
\begin{equation*}
\begin{split}
    [P_{j_1}P_{j_2}...P_{j_t}, H] &= [P_{j_1}P_{j_2}...P_{j_t}, h_R]\\
    &= \sum_{\{p: \sigma_p \in h_R\}} \gamma_p [P_{j_1}P_{j_2}...P_{j_t}, \sigma_p]\\
\end{split}
\end{equation*}
for some $R$. The number of Pauli words included in $h_R$ is upper bounded by $4^k$, and thus overall we need to compute a maximum of 
\begin{equation*}
    {k \choose t} |E| 4^k
\end{equation*}
terms. When $t = \left\lfloor{\frac{k}{2}}\right \rfloor$, this reaches a maximum of:
\begin{equation*}
    \frac{k!}{\left\lfloor{\frac{k}{2}}\right \rfloor! \left\lceil{\frac{k}{2}}\right \rceil!}|E| 4^k 
\end{equation*}
If we consider the limit as $k$ goes to infinity, we can use Stirling's approximation,
\begin{equation}
    {n \choose k} \approx \sqrt{\frac{n}{2 \pi k(n-k)}}\frac{n^n}{k^k(n-k)^{n-k}}
\end{equation}
to write: 
\begin{equation*}
    \begin{split}
        \frac{k!}{\left\lfloor{\frac{k}{2}}\right \rfloor! \left\lceil{\frac{k}{2}}\right \rceil!}|E| 4^k &\approx \sqrt{\frac{2}{\pi  k}}|E|4^{k+1}
    \end{split}
\end{equation*}
In the case of chemistry, we can to encode the Hamiltonian so that the locality $k$ scales as $\log(n)$, so we can write: 
\begin{equation*}
    \begin{split}
         \sqrt{\frac{2}{\pi  k}}|E|4^{k+1} &=  O(|E|n \log(n)^{-1/2})
    \end{split}
\end{equation*}
Thus, the scaling is sub-exponential in system size.
\end{document}